\documentclass[a4paper,11pt]{article}
\usepackage{pos}

\title{Measurement of $\mathrm{D}$-meson production as a function of charged-particle multiplicity in proton--proton collisions at $\sqrt{s} = 13$ TeV with ALICE at the LHC}

\ShortTitle{D-meson production in pp collisions at $\sqrt{s} = 13$ TeV with ALICE}
\manuallySeparateAuthors

\author*[a]{Yoshini Bailung}
\author{for the ALICE Collaboration}

\affiliation[a]{Department of Physics, Indian Institute of Technology Indore \\
Indore, Madhya Pradesh, India}

\emailAdd{yoshini.bailung@cern.ch}

\abstract{Heavy quarks (charm and beauty) are produced in hard-scattering processes and the study of their production in proton--proton (pp) collisions is an important test for calculations based on perturbative Quantum Chromodynamics (pQCD). Heavy-flavor production as a function of charged-particle multiplicity provides insight into the processes occurring at the partonic level and the interplay between the hard and soft particle production mechanisms in pp collisions.

In this contribution, measurements of open heavy-flavor production as a function of multiplicity, via the study of the $\mathrm{D}$-meson self-normalized yields in pp collisions at the center-of-mass energy of $\sqrt{s} = 13$ TeV is presented. The $\mathrm{D}$-meson self-normalized yield is found to increase stronger than linearly with increasing charged-particle multiplicity. The measurements are compared to theoretical model calculations, and with the results at $\sqrt{s} = 7$ TeV.}

\FullConference{%
 
 The Ninth Annual Conference on Large Hadron Collider Physics - LHCP2021\\
 7-12 June 2021\\
 Online
}

\begin{document}
\maketitle
\section{Introduction}
The study of heavy-flavor hadron production in pp collisions at the LHC provides a stringent test to pQCD calculations. Due to their large masses, the heavy quarks are produced via hard scatterings and studied through the production of open heavy-flavor hadrons. Measurements of open charm production in pp collisions as a function of charged-particle multiplicity provide information on the interplay between hard and soft processes in particle production. At high energies, multi-parton interactions (MPI) can have a major contribution in particle production and ultimately the total event multiplicity~\cite{Sjostrand:2004pf,Sjostrand:1987su,Weber:2018ddv}. In this contribution, production of D mesons ($\rm{D^0,D^+,D^{*+}}$) as a function of charged-particle multiplicity in pp collisions at $\sqrt{s} = 13$ TeV at mid-rapidity is studied. A similar study was carried out in pp collisions at $\sqrt{s} = 7$ TeV~\cite{ALICE:2015ikl}, where a stronger than linear trend of the D-meson yields with relative charged-particle multiplicity was observed. In this contribution, the new results of D-meson production in pp collisions performed with ALICE~\cite{ALICE:2008ngc} using the data collected during Run 2 of the LHC at $\sqrt{s} = 13$ TeV will be presented.

\section{Analysis Strategy}

Multiplicity is defined as the number of track segments or $`$tracklets' ($N_{\rm tracklets}$) reconstructed from hits in the two layers of the Silicon-Pixel Detector (SPD) of the Inner Tracking System (ITS) within $|\eta|<1$. In order to transform the multiplicity definition based on the tracklets to a physical quantity, a conversion is done from $N_{\rm tracklets}$ to the number of charged particles ($N_{\rm ch}$). It is carried out using the correlation map between $N_{\rm tracklets}$ and $N_{\rm ch}$ obtained by Monte Carlo simulations, to which a first order polynomial fit is applied. The slope and offset parameters of the fit are used to convert the $N_{\rm tracklets}$ to charged-particle multiplicity density ($\rm{d} \it{N}_{\rm ch}/d\eta$) intervals . The ITS is further used for secondary vertex reconstruction and tracking. The Time Projection Chamber (TPC) is used for tracking and particle identification (PID), and the Time of Flight Detector (TOF) is used for PID.

The D-meson raw yields are reconstructed at mid-rapidity from their hadronic decay channels, $\rm{D^0 \rightarrow K^- \pi^+}$ with a branching ratio (BR) of (3.87 ± 0.05)\%, $\rm{D^+ \rightarrow K^- \pi^+ \pi^+}$ with BR = (9.13 ± 0.19)\%, and $\rm{D^{*+} \rightarrow D^0\pi^+}$ with BR = (67.7 ± 0.05)\%. PID and topological selections are applied to reduce the combinatorial background of D mesons reconstructed from their decay particle tracks. The raw yields are extracted by fitting the invariant-mass distributions. The raw yields are then corrected by the acceptance and efficiency factors, trigger efficiencies, and the number of events for every multiplicity interval. The yields in each multiplicity interval ($\rm{Y^{mult}}$) are presented relative to those in the multiplicity integrated sample ($\rm{Y^{mult\ int}}$). The D-meson self-normalized yields are constructed as 
\begin{equation}
\rm{Y_{corr}^{mult}} = \left(\frac{Y^{mult}}{((Acc\times \epsilon_{prompt}^{mult})\times \it{N}\rm_{event}^{mult})/\epsilon^{trg}_{mult}}\right) \Biggm/ \left(\frac{Y^{mult\ int}}{((Acc \times \epsilon_{prompt}^{mult\ int})\times \it{N}\rm_{event}^{mult\ int})/\epsilon^{trg}_{mult\ int}}\right)
\label{selfnormalizedyield}
\end{equation}
where, in the numerator $\rm{Y^{mult}}$ is the extracted raw yield, $\rm{Acc\times \epsilon_{prompt}^{mult}}$ is the acceptance times efficiency factor for D mesons originating from the collision (prompt), $\rm{\it{N}\rm^{mult}_{event}}$ is the number of events, and $\rm{\epsilon^{trg}_{mult}}$ is the trigger efficiency for a particular multiplicity interval. The numerator is normalized to the corresponding quantity for the multiplicity integrated sample. The values of $\rm{d} \it{N}_{\rm{ch}}/d\eta$ are normalized with the mean charged-particle multiplicity density ($\langle\rm{d} \it{N}_{\rm{ch}}/d\eta\rangle$). 

\section{Results}

\begin{figure}[!ht]
    \centering
    \includegraphics[scale = 0.49]{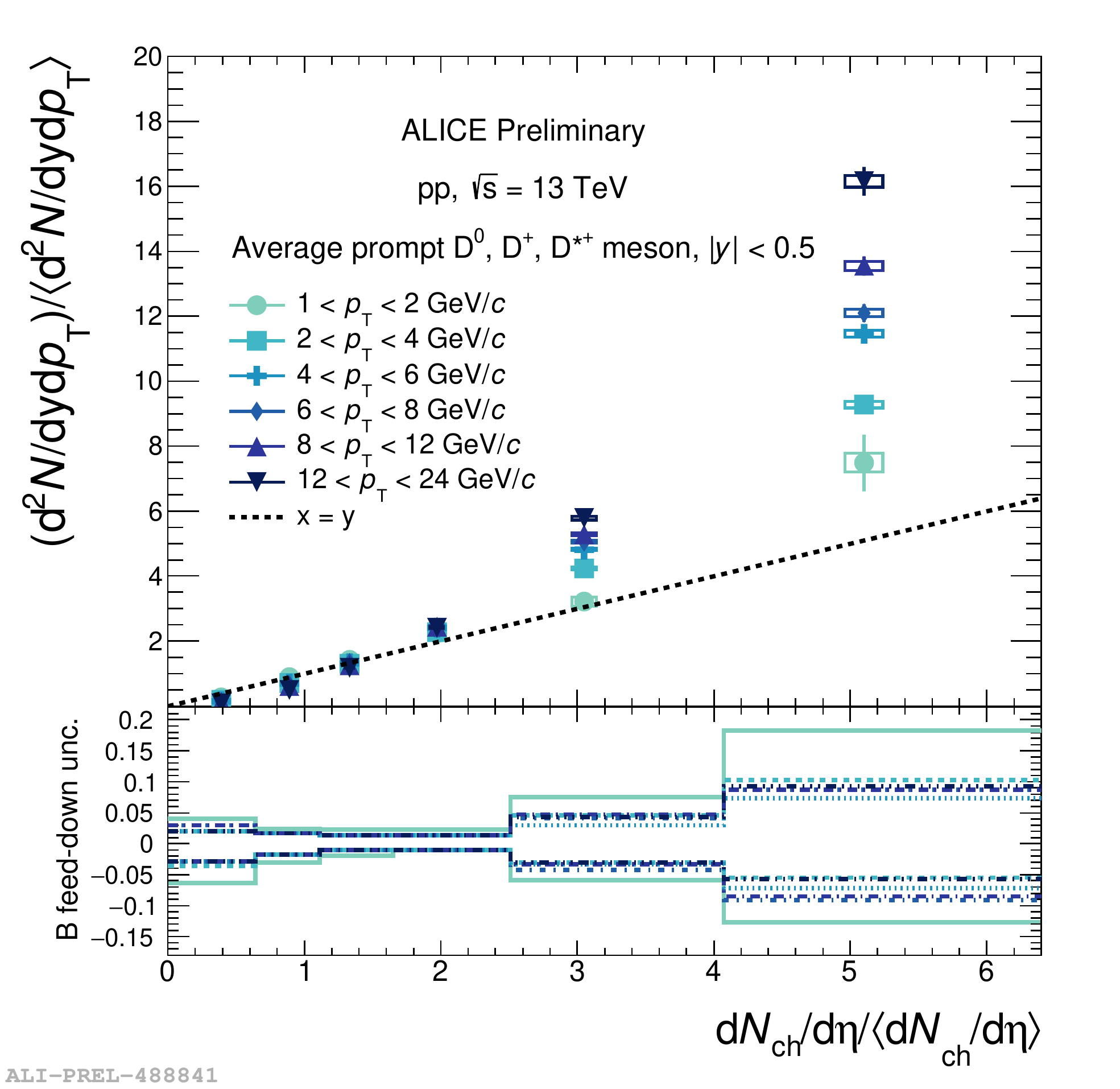}
    \caption{Average of prompt $\rm{D^{0},D^{+}\ and \ D^{*+}}$ self-normalized yields as a function of the relative charged-particle multiplicity at mid-rapidity in various $\it{p}_{\rm{T}}$ intervals. The results are presented in the top panel with statistical (vertical bars) and systematic (boxes) uncertainties. The uncertainty associated with the contribution fraction from beauty hadron decays is drawn in the bottom panels.}
    \label{fig:f1}
\end{figure}

\begin{figure}[!ht]
    \flushleft
    \includegraphics[scale = 0.375]{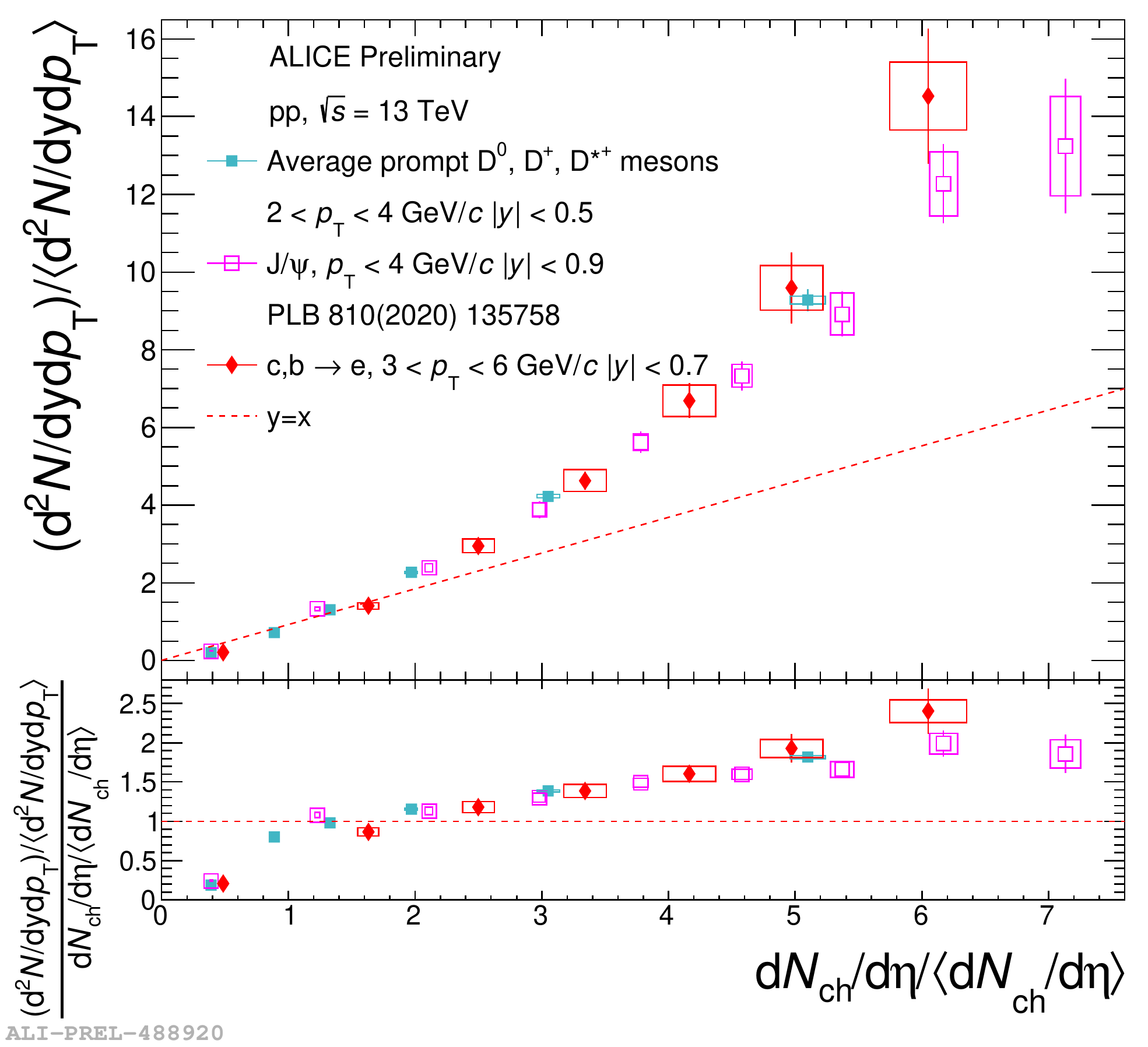}
    \includegraphics[scale = 0.375]{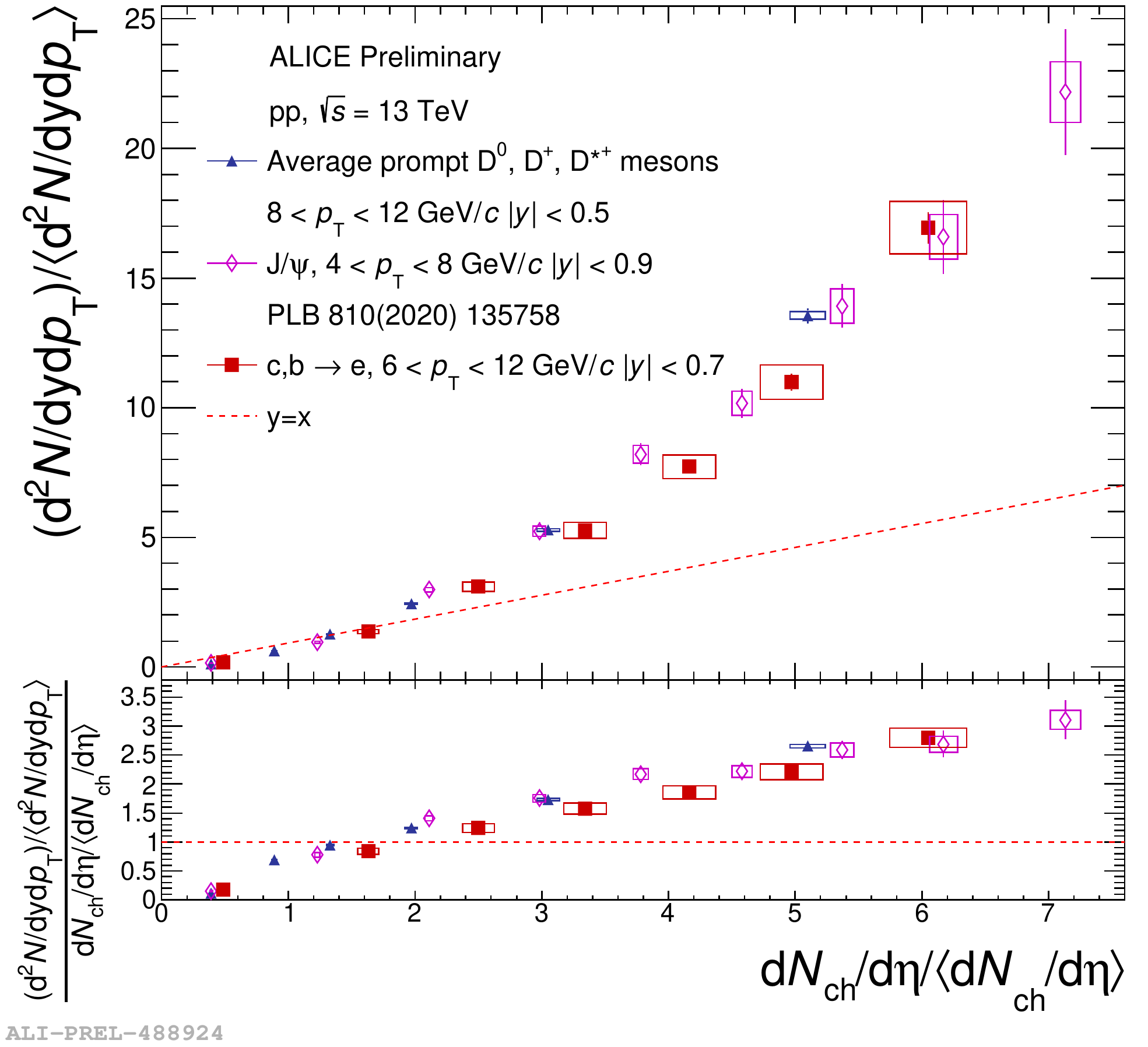}
    \caption{Average D-meson, J/$\psi$, and heavy-flavor decay electrons self-normalized yields as a function of relative charged-particle multiplicity at mid-rapidity in pp collisions at $\sqrt{s} = 13$ TeV at compatible low (left) and high (right) $\it{p}_{\rm{T}}$ intervals. The bottom panel shows the comparison of double ratios between the species as a function of relative charged-particle multiplicity.}
    \label{fig:f2}
\end{figure}

The average prompt D-meson self-normalized yields are measured at mid-rapidity ($|\it{y}|<$ 0.5) as a function of charged-particle multiplicity in pp collisions at $\sqrt{s} = 13$ TeV as shown in Fig.~\ref{fig:f1}. The results show a stronger than linear increase of D-meson yields with increasing charged-particle multiplicity, and a steeper increase of the yields with increasing $\it{p}_{\rm{T}}$. The new results provide improved precision of the measurements at high multiplicities and a good agreement with the results at $\sqrt{s} = 7$ TeV. In Fig.~\ref{fig:f2}, comparisons with different heavy-flavor species in similar $\it{p}_{\rm{T}}$ intervals show compatibility with the faster than linear increase with multiplicity and the strong $\it{p}_{\rm{T}}$ dependence. Average D-meson measurements are compatible with $\rm{J}/\psi$~\cite{ALICE:2020msa} and electrons from heavy-flavor hadron decays (c,b$\rightarrow$e) in pp collisions at $\sqrt{s} = 13$ TeV.

\begin{figure}[!ht]
    \centering
    \includegraphics[scale = 0.56]{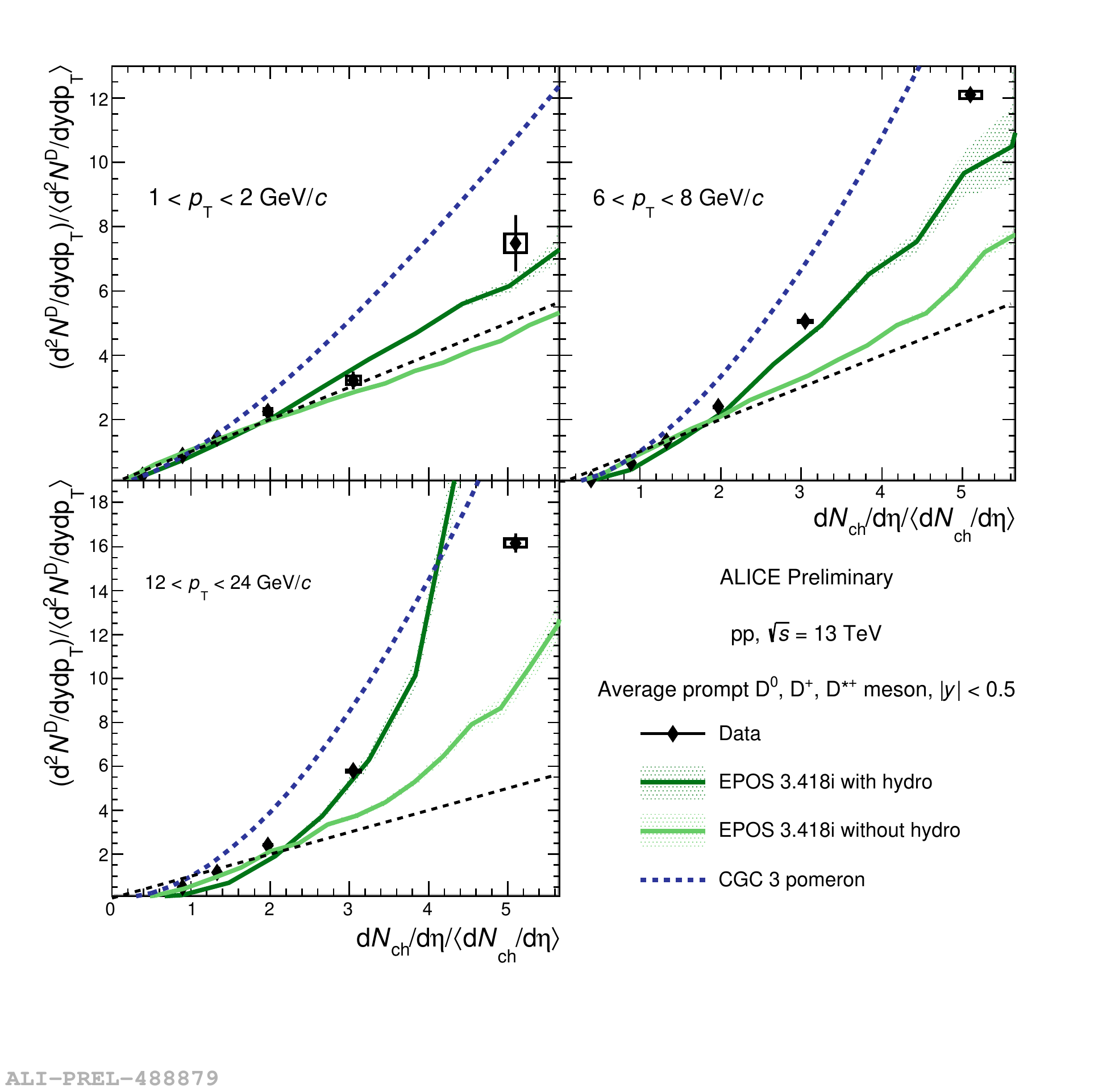}
    \caption{The average prompt $\rm{D^{0},D^{+}\ and \ D^{*+}}$ self-normalized yields vs relative charged-particle multiplicity in pp collisions at $\sqrt{s} = 13$ TeV at mid-rapidity is compared with model predictions in different $\it{p}_{\rm{T}}$ intervals.}
    \label{fig:f3}
\end{figure}

In Fig.~\ref{fig:f3}, comparisons with model predictions from EPOS3~\cite{Werner:2013tya} and 3-pomeron Color Glass Condensate (CGC)~\cite{Schmidt:2020fgn} are shown. EPOS3 with a hydrodynamic evolution shows a good agreement with the faster than linear trend of D-meson yields at low and intermediate multiplicity, however overestimating at large multiplicity values. EPOS3 without a hydrodynamic component predicts a small increase in the D-meson production, underestimating the obtained results. The hydrodynamic phase reduces the number of charged particles produced in EPOS, leading to the differences between the two modes. The 3-pomeron exchange CGC model reproduces the faster than linear trend of the D mesons, although overestimates the results. 


\section{Summary}

In this contribution, the new measurements of D-meson production as a function of charged-particle multiplicity in pp collisions at $\sqrt{s} = 13$ TeV are presented. The D-meson yields show a faster than linear increase with increasing multiplicity. 
The results show good agreement with the D-meson measurements in pp collisions at $\sqrt{s} = 7$ TeV, and J/$\psi$ and electrons from heavy-flavor measurements in pp collisions at $\sqrt{s} = 13$ TeV. EPOS3 with a hydrodynamic evolution describes the data. However, EPOS3 without any hydrodynamics and 3-pomeron CGC are seen to underestimate and overestimate the data, respectively. These measurements will further improve in Run 3, with higher luminosity and improved detector performance~\cite{ALICE:2013nwm}.

\end{document}